\documentclass[aps,prc,notitlepage]{revtex4-1}

\usepackage{graphicx}
\usepackage{soul}
\usepackage[normalem]{ulem}
\usepackage{color}
\definecolor{purple}{rgb}{0.75,0,0.75}

\newcommand{\beq}{\begin{equation}}
\newcommand{\eeq}{\end{equation}}
\newcommand{\bea}{\begin{eqnarray}}
\newcommand{\eea}{\end{eqnarray}}
\newcommand{\nn}{\nonumber}

\begin{document}

\title{Hypernuclei and the hyperon problem in neutron stars}
\author{ Paulo F. Bedaque }
\affiliation{Department of Physics, University of Maryland, 
College Park, Maryland 20742, USA}
\author{Andrew W. Steiner}
\affiliation{
Institute for Nuclear Theory, University of Washington,
Seattle, Washington 98195, USA \\
Department of Physics and Astronomy, University of
Tennessee, Knoxville, Tennessee 37996, USA \\
Physics Division, Oak Ridge National Laboratory, Oak
Ridge, Tennessee 37831, USA}

\begin{abstract}
The likely presence of $\Lambda$ baryons in dense hadronic matter
tends to soften the equation of state to an extend that the observed
heaviest neutron stars are difficult to explain. We analyze this
``hyperon problem'' with a phenomenological approach. First, we review
what can be learned about the interaction of $\Lambda$ particle with
dense matter from the observed hypernuclei and extend this
phenomenological analysis to asymmetric matter. We add to this the
current knowledge on non-strange dense matter, including its uncertainties,
to conclude that the interaction between $\Lambda$s and dense
matter has to become repulsive at densities below three times 
the nuclear saturation density.
\end{abstract}

\maketitle

\section{Introduction}

One of the main motivations for the study of matter at densities in
excess of the nuclear saturation density is, besides the application
to the physics of neutron stars, the possibility of unveiling new
phases of matter. Among these more exotic phases, quark matter is the
most sought out as its existence is all but guaranteed by the fact
that QCD becomes weakly coupled at arbitrarily high densities.
However, even if the transition to quark matter occurs at densities
inaccessible to neutron stars, nucleons are not obviously the only
relevant degrees of freedom. Of particular interest is the possible
existence of $\Lambda$ particles. Due to a combination of
circumstances they are likely the first one to appear as the density
of matter increases. First, they are the lightest baryon (besides
nucleons). Second, they are neutral so their appearance does not incur
in the appearance of an electron and the consequent cost of an
electron Fermi energy. Finally, as we will argue below, the
phenomenology of hypernuclei strongly suggests that $\Lambda$s are
attracted to neutron matter, lowering the energetic cost of $\Lambda$s
even further.

How can the presence of $\Lambda$s be inferred from neutron star
observations? For a given (zero temperature) equation of state,
general relativity predicts a specific relation between the star mass
and radius. Radii are very difficult to measure because of the large
systematic uncertainties involved~\cite{SLB13,Lattimer14b}. Current
constraints on the mass-radius relation from radius measurements are
not strong enough to significantly constrain the presence of
$\Lambda$s. However, one feature of general relativity is particularly
helpful for putting bounds on the equation of state. For a given
equation of state there is a maximum mass beyond which the star will
collapse into a black hole, regardless of its radius. Thus, the
discovery of two stars with masses around two solar masses requires
fairly stiff equations of state and is in tension with the presence of
$\Lambda$ particles, which soften the equation of state significantly.
In fact, the equation of state of matter formed by nucleons only can
be softened by having neutron on the top of the Fermi sphere
transforming (through weak interactions) into $\Lambda$s at rest. As a
result, the same baryon density can be achieved with a smaller energy
density. A rough estimate of the density for the onset of $\Lambda$
appearance can be obtained by equating the neutron Fermi energy to the
mass difference between neutrons and $\Lambda$s. For realistic
equations of state this density is around a few times nuclear
saturation density, well inside the range relevant for neutron stars.
Simple calculations assuming a weak interaction between neutrons and
$\Lambda$s show that the softening of the equation of state makes it
very difficult for an equation of state with $\Lambda$ degrees of
freedom to support a star with a mass around two solar masses as
recently observed~\cite{Demorest:2010bx,Antoniadis:2013pzd}. This
apparent contradiction is sometimes referred to as ``the hyperon
problem''~\cite{Chen11,Weissenborn12,Dexheimer13,Whittenbury14,Yamamoto14,Drago14,Lopes14,Lonardoni14,vanDalen14}.
A few ways to solve this paradox immediately come to mind. Simply
assuming a harder equation of state for nucleonic matter does not
necessarily solve the problem and can actually make it worse. A hard
equation of state for the neutrons {\it lowers} the density at which
hyperons appear. Another solution would be to assume that the
interaction between $\Lambda$s and neutrons is sufficiently repulsive
to raise the effective mass of the $\Lambda$ particle in dense neutron
matter, raising the threshold for $\Lambda$ appearance and postponing
the softening of the equation of state to irrelevant densities.
However, we know that $\Lambda$s are in fact {\it attracted} to
nuclear matter since stable (against strong interactions) bound states
of a $\Lambda$ particle with a variety of nuclei are know (for a
review see~\cite{Hashimoto2006564}). Microscopic meson exchange models
of $\Lambda$-nucleon interaction also indicates an attraction between
$\Lambda$ and neutrons. They are not sufficient to explain away the
``hyperon problem'' but the addition of a repulsive enough three-body
force ($\Lambda$NN) might. In fact, one such a model has been
constructed~\cite{Lonardoni14}. The difficulty with this approach is
that the $\Lambda$NN three-body force is very little constrained by
either experiment or theory. A lattice QCD calculation of the
$\Lambda$N and $\Lambda$NN interaction, although very difficult, is
likely to come out in the near future and, in fact, might be the first
reliable calculation of forces between baryon to be accomplished
\cite{Beane:2003yx,Beane:2006gf,Beane:2012ey}.
  
In this paper we will not discuss any microscopic model and, instead,
take a radically phenomenological approach. First we will review --
and slightly extend -- what can be learned about the properties of
$\Lambda$s in a dense medium from a simple-minded phenomenology of
hypernuclei. More precisely, we will review the extraction of the
$\Lambda$ mass shift in {\it nuclear} matter and discuss a bound on
the $\Lambda$ mass shift in {\it neutron} matter. This analysis will
tell us about the $\Lambda$ properties in neutron/nuclear matter at
nuclear saturation densities. We will then use the existence of the
neutron stars with $M\approx 2 \mathrm{M}_{\odot}$ to establish constraints on
the change of these properties with density. We will show that, to no
surprise, the attraction between $\Lambda$s and neutrons at nuclear
densities has to quickly into a repulsion and will quantify this
statement. We will end by commenting on microscopic mechanisms for
this change as well as possible hypernuclei experiments which could
help our approach narrow down the range of empirically acceptable
equations of state.

\section{Hypernuclei and the interaction between $\Lambda$ and 
nuclear/neutron matter}

\begin{figure*}
\includegraphics[width=10cm]{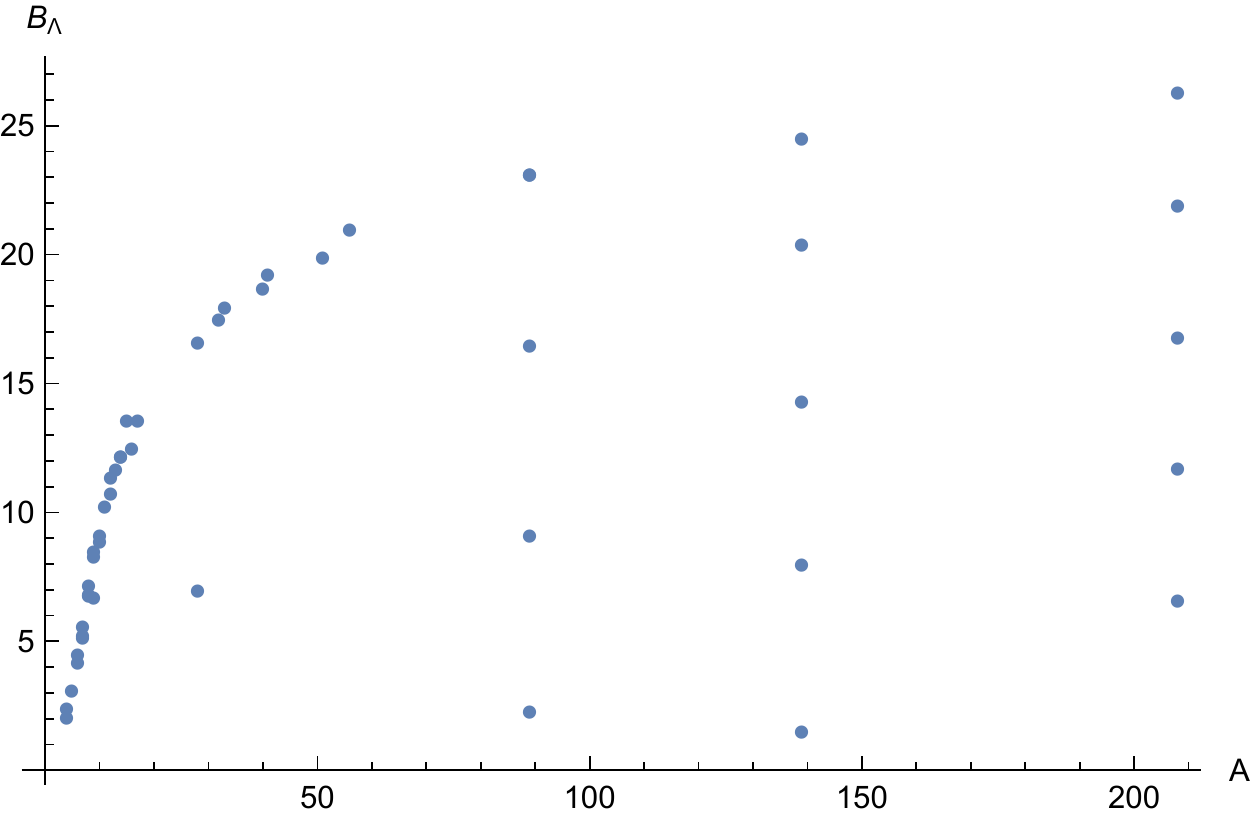}
\caption{$\Lambda$ binding energies of known hypernuclei (in MeV).
  Data taken from~\cite{Samanta:2005kd} and~\cite{Hashimoto2006564}.}
\label{fig:hypernuclei}
\end{figure*}

The existence of bound states of one $\Lambda$ particle to a number of
nuclei indicates that the interaction between $\Lambda$s and nucleons
is mostly attractive. A more quantitative statement statement can be
made if we consider the binding energies is some detail.
Fig.~\ref{fig:hypernuclei} shows the measured binding energies of a
$\Lambda$ particle as a function of the mass number $A$ of the nucleus
(from the data compiled in~\cite{Samanta:2005kd,Hashimoto2006564}). In
some cases, more than one hypernucleus with the same value of $A$
appears. In the case of small $A$ they correspond to nuclei with
different atomic numbers $Z$. For the larger values of $A$ they
correspond to different excited states of the same nucleus
corresponding to different values of the angular momentum $l$ as, for
instance, the five states in ${}^{208}$Pb.

The detailed value of these binding energies can only be computed in
detailed calculations, either starting from the largely unknown
$\Lambda$-nucleon and $\Lambda$-nucleon-nucleon forces or, more
ambitiously, from QCD. Both approaches are in their infancy but, once
they succeed one can imagine extend them to the study of hyperon in a
neutron medium. However, simple phenomenological methods, akin to the
mass formula for non-strange nuclei, can shed  light on the
data and have the additional advantage of being immediately extendable
to strange dense matter. The basic observation is that the interaction
between one $\Lambda$ and the nucleons is short range; in fact, it is
expected to be of even shorter range than nuclear forces on account of
the absence of the one-pion exchange component which cannot occur for
an isoscalar particle. Thus, a $\Lambda$ particle inside a
hypernucleus interacts with only a few nucleons around it. Since
nuclei have a fairly constant density, the main effect of the
$\Lambda$-nucleon interaction is to provide a spherical constant
potential well inside which the $\Lambda$ particle moves freely. The
depth of this well is also expected to be the same for every nucleus,
again on account of nuclear saturation. This simple model corresponds
to a $\Lambda$ binding energy given by energy of a $\Lambda$ particle
in a spherical well with radius proportional to $A^{1/3}$ plus a fixed
shift:
\beq\label{eq:two_param}
B_\Lambda = \Delta E - c \frac{z^2_{l n}}{A^{2/3}},
\eeq 
where $z_{l n}$ is the $n^{th}$ zero of the $l^{th}$ spherical Bessel
function, $A$ is the mass number of the nucleus {\it not} including
the $\Lambda$ and $\Delta E$ and $c$ are fitting parameters. The
values $\Delta E= 24.6$ MeV and $c=68.7$ MeV were obtained from a fit
of the $A\geq 8$ nuclei (we exclude very small nuclei where this
approach clearly does not make sense). This simple fit does a
reasonable job at the qualitative level but leaves a lot of room for
improvement. As a way of describing the goodness of the fit we notice
that, if we add (in quadratures) a theoretical uncertainty of $10\%$,
the $\chi^2$ per degree of freedom of this fit is about 10.

\begin{figure}
\includegraphics[width=8cm]{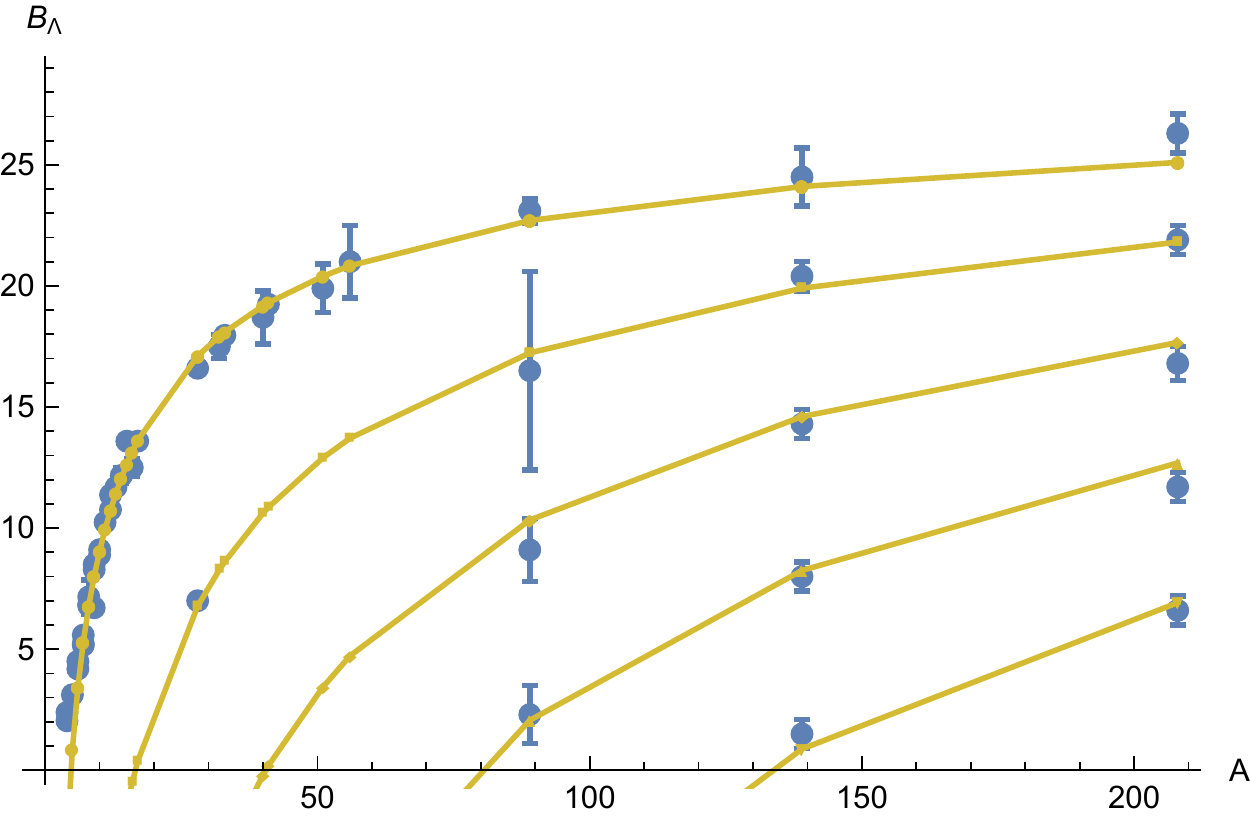}
\caption{$\Lambda$ binding energies of known single-$\Lambda$
  hypernuclei and the three parameter fit from
  eq.~\ref{eq:three_param}. Only nuclei with $A\geq 8$ were used in
  the fit. }
\label{fig:hypernuclei_c1c2e0_a}
\includegraphics[width=8cm]{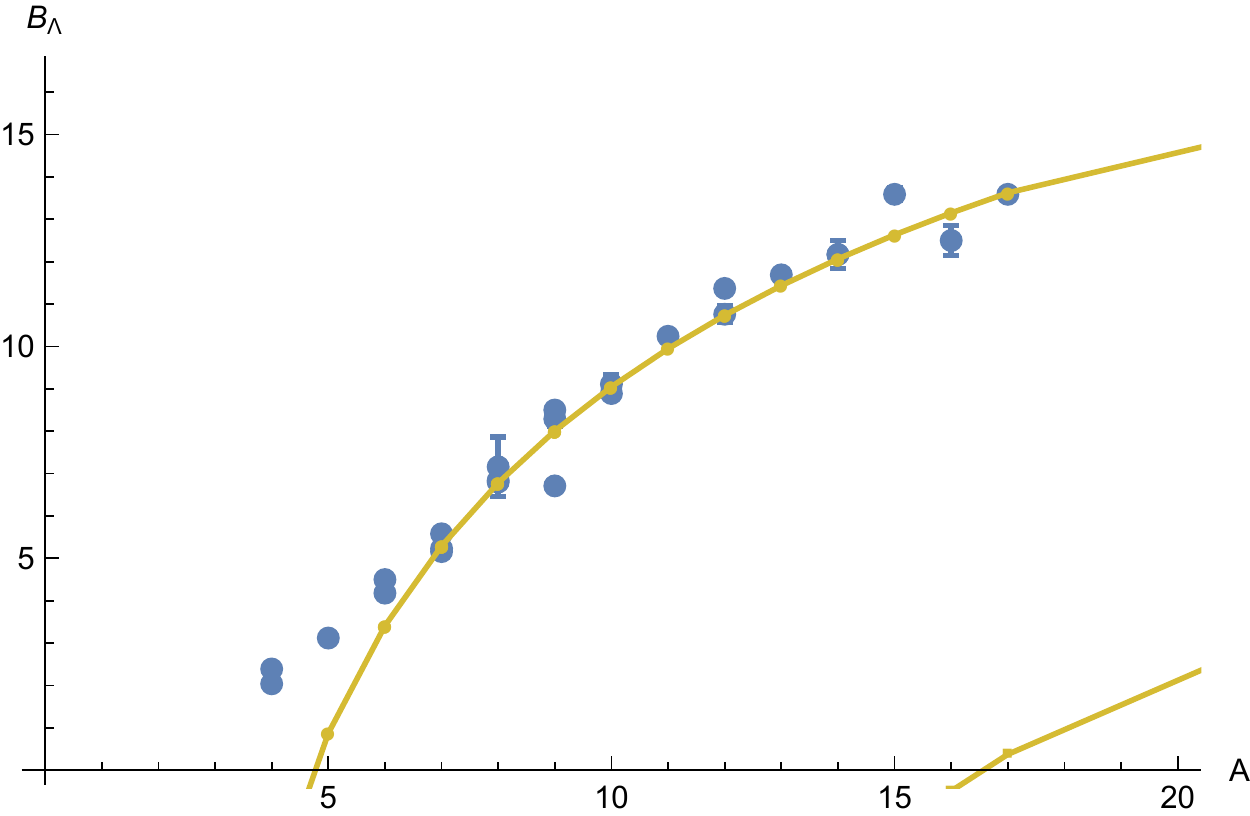}
\caption{Detail of fig.~\ref{fig:hypernuclei_c1c2e0_a} for low A.}
\label{fig:hypernuclei_c1c2e0_b}
\end{figure}

Two improvements on the model in eq.~\ref{eq:two_param} make it work
much better. One is to use a more precise description of nuclear radii
in the kinetic energy term. Nuclear radii, as measured in elastic
electron scattering, can be parametrized by $R=r_0 (A^{1/3} +1.565
A^{-1/3} - 1.043 A^{-1})$\cite{Friedrich1982192}. The parameter $r_0$
is absorbed in $c$ and is unimportant. This parametrization of the
nuclear radius provides a more accurate determination of the kinetic
energy of the $\Lambda$ as a function of $A$. An alternative procedure
would be to let the constants 1.565 and $-$1.043 to float during the fit
but similar results are obtained so, for the sake of brevity, we will
not pursue it here. At the boundary of the nucleus the $\Lambda$
particle potential has to change to zero. Using a smooth shape for the
potential as a function of the distance from the center (as opposed to
the step function implicit in eq.~\ref{eq:two_param}) might be more
realistic. A similar effect is caused by the thin neutron skin near
the boundary. One way of dealing with these corrections is to solve
for the $\Lambda$ energy levels in a Woods-Saxon potential as done in
\cite{PhysRevC.38.2700}. In order to obtain analytic expressions, we
chose to include these effects perturbatively, assuming only that the
distance over which the potential transitioned to zero was much
smaller than the nucleus radius. A simple first order perturbative
calculation leads to a dependence of the energy of the form $\sim
1/R^3 \sim 1/A$ for this contribution. These two effects combined
change eq.~\ref{eq:two_param} to
\beq
\label{eq:three_param}
B_\Lambda = \Delta E - c \frac{z^2_{l n}}
{ (A^{1/3} +1.565 A^{-1/3} -  1.043 A^{-1})^2  } + \frac{e}{A}.
\eeq 
Fitting the three parameters $\Delta E, c$ and $e$ in
eq.~\ref{eq:four_param} to the binding energy of the $\Lambda$ to
nuclei with $A\geq 16$ we obtain $\Delta E = 28.5$ MeV, $c=119$ MeV
and $e=-64.7$ MeV. We again add in quadrature a theoretical uncertainty
of $10\%$ to the experimental uncertainty. The plots in
fig.~\ref{fig:hypernuclei_c1c2e0_a} and
fig.~\ref{fig:hypernuclei_c1c2e0_b} show this fit compared to the
available data. This fit works very well for most nuclei with $A\geq
8$, including the excited levels of ${}^{208}$Pb. By varying details
of the fit, as the minimum value of $A$ and/or the assumed theoretical
error a likely range for $27.5 {\rm \ MeV} \leq \Delta E \leq 29.5
{\rm \ MeV}$ is obtained.

These results indicate that the binding energy of a $\Lambda$ particle
in nuclear matter is about $28$ MeV. But in neutron stars, the more
relevant quantity is the shift in energy of a $\Lambda$ particle in
{\it neutron} matter. The data set, however, includes hypernuclei with
neutron excess $(A-2Z)/A$ varying from $-$0.07 to 0.21 among larger
nuclei with $A\geq 10$ and it is a natural question whether this
spread in neutron excess can be explored in order bound the energy
shift in neutron matter.

Starting from the good fit in eq.~\ref{eq:three_param} (very similar
to the one in~\cite{PhysRevC.38.2700}) we can address the variation of
$\Lambda$ mass with the neutron excess by adding an extra term
proportional to the neutron excess $(A-2Z)/A$ to
eq.~\ref{eq:three_param}:

\beq
\label{eq:four_param}
B_\Lambda = \Delta E - c \frac{z^2_{l n}}
{   (A^{1/3} +1.565 A^{-1/3} -  1.043 A^{-1})^2  } + 
\frac{e}{A}+d \left(\frac{A-2Z}{A}\right)^2.
\eeq 

\begin{figure}[!tbp]
\includegraphics[width=8cm]{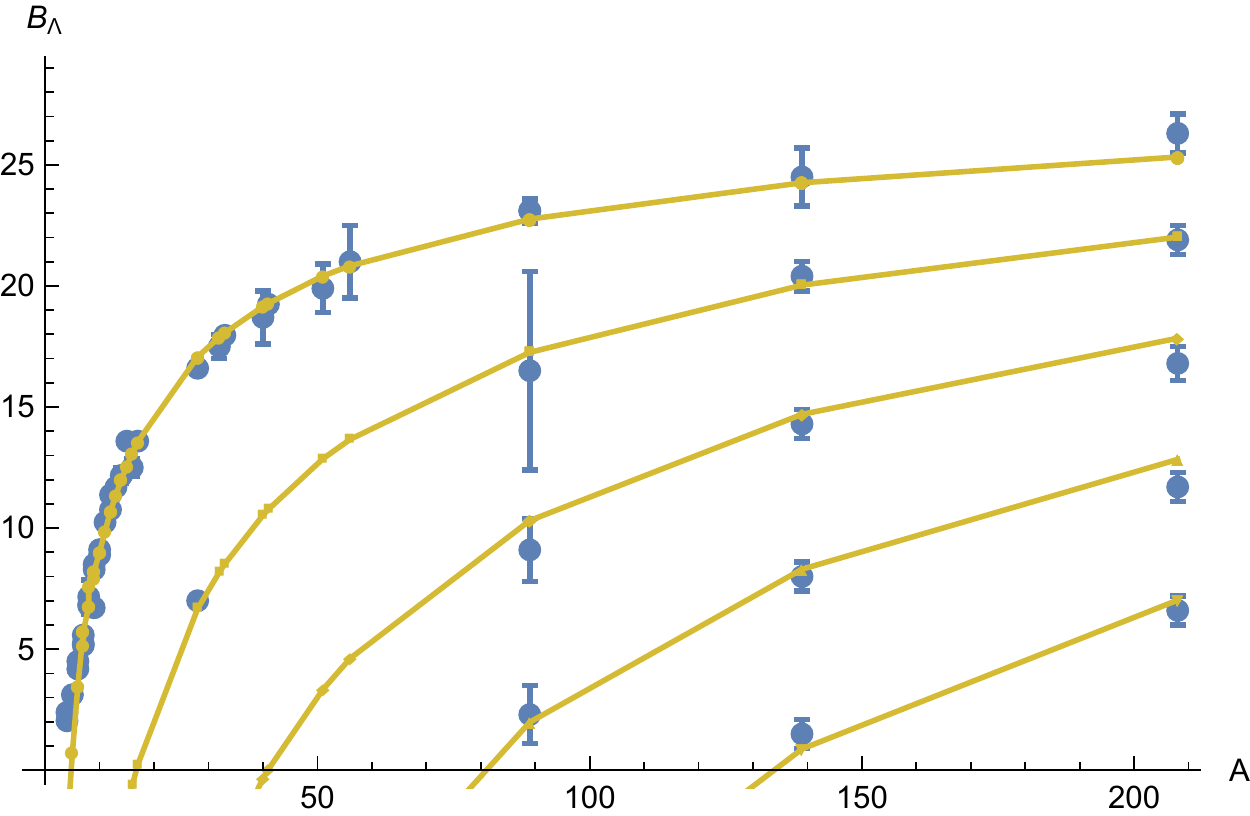}
\caption{$\Lambda$ binding energies of known single-$\Lambda$
  hypernuclei and the four parameter fit from
  eq.~\ref{eq:four_param}. }
\label{fig:hypernuclei_2_a}
\includegraphics[width=8cm]{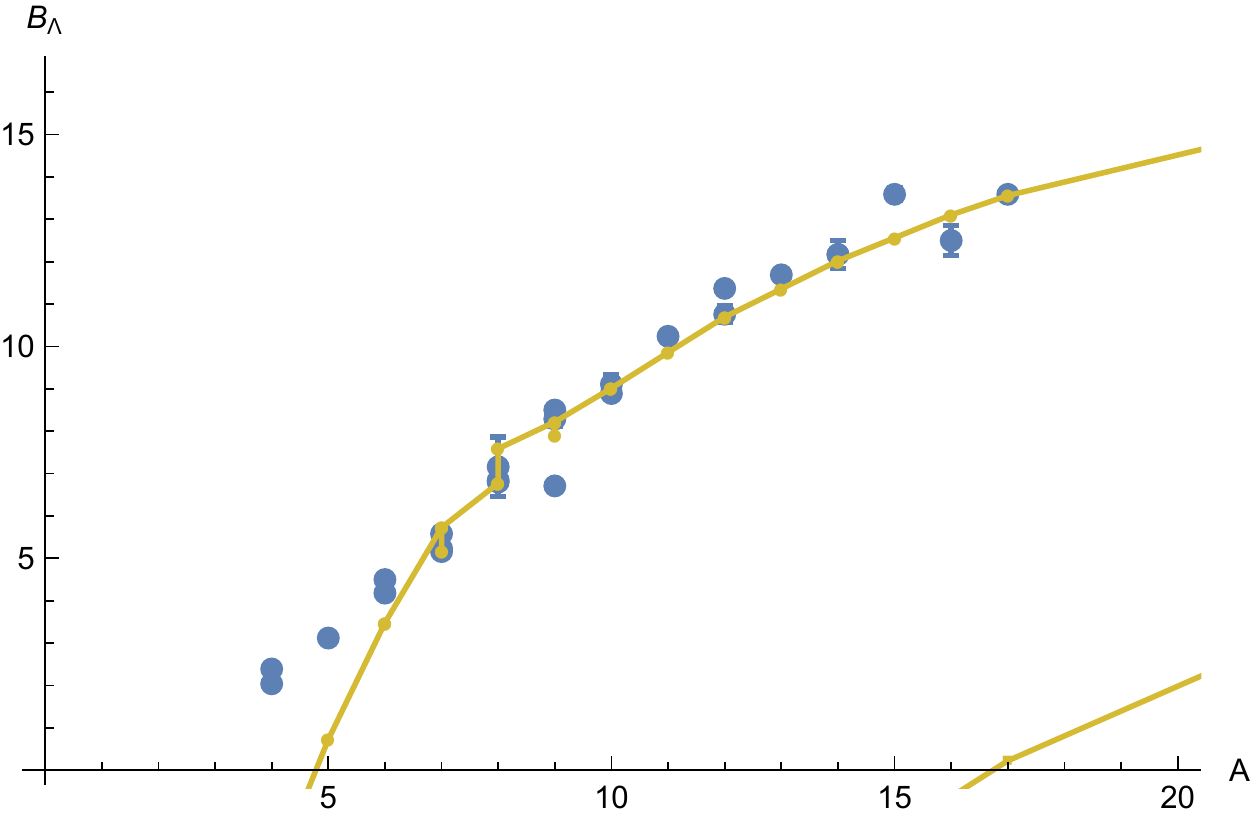} 
\caption{Detail from fig.~\ref{fig:hypernuclei_2_a} for low A. 
Only nuclei with $A\geq8$ were used in the fit. }
\label{fig:hypernuclei_2_b}
\end{figure}

Terms linear in the neutron excess $(A-2Z)/A$ are not expected as they
are proportional to isospin breaking terms. In fact, a term
proportional to $(A-2Z)/A$ does not improve the fit. A four parameter
fit of the $A\geq 8$ nuclei gives $\Delta E = 28.5$ MeV, $c=120$ MeV,
$e=-65.1$ MeV and $d=4.99$ MeV and is compared to the data in
fig.~\ref{fig:hypernuclei_2_a} and \ref{fig:hypernuclei_2_b}. This fit
is modestly better than the simpler three-parameter fit in
eq.~\ref{eq:three_param} suggesting a small dependence of the binding
energies on the neutron excess. In order to estimate the range of
acceptable values of $d$ we computed the $\chi^2$ per degree of
freedom of a fit of eq.~\ref{eq:four_param} with fixed value of $d$.
We find that it changes from $\chi^2=1.3$ at $d=4.99$ MeV (the
preferred value) to $\chi^2=2$ at $d=20$ MeV or $d=-10$ MeV. Changes
in the estimated theoretical uncertainty and minimum value of $A$ used
in the fit do not change this result by more than its uncertainty. We
take this as evidence that values of $d$ outside the range $-10{\rm
  \ MeV} \alt d \alt15 {\rm \ MeV}$ are disfavored by hypernuclei
data. It should be stressed however, that the systematic errors
involved in the arbitrary choice of parametrizations are difficult to
estimate and are not included in a rigorous way in our estimate. On
the experimental side, the observation of further large, neutron rich
hypernuclei would help constrain the value of the parameter $d$.

The small dependence of the $\Lambda$ binding energy on the neutron
excess is expected, at least at small enough densities. In fact, the
shift in the energy of a $\Lambda$ particle at small densities is
proportional to the density of the particles in the medium (and
proportional to the forward scattering
amplitude~\cite{Eletsky:1996jg}). But, due to isospin symmetry, the
scattering amplitude for $\Lambda$-proton and $\Lambda$-neutron is
(approximately) the same. Whether the $\Lambda$ scatters out of a
density $n$ of neutrons or a density $x n$ of protons and $(1-x)n$ of
neutrons, the total shift in energy is the same. At high enough
densities, the dependence of the energy shift with the density of
scatterers is no longer linear and a dependence on the proton fraction
appears, even if isospin symmetry were exact.

\section{Equation of state including hyperons}

We can now use the lessons from the previous section, in special the
estimated shift in energy of a single $\Lambda$ particle in neutron
matter, to construct equations of state for matter at the densities
relevant for neutron star physics. The first observation is that the
presence of too many $\Lambda$s softens the equation of state too much
to accommodate two solar masses stars. As we will see below,
typically, a
$\Lambda$ fraction of no more than about $10\%$ is required.
Consequently, the $\Lambda-\Lambda$ interaction plays a small role and
will be neglected. The energy density is then a linear function of the
$\Lambda$ fraction $y$. We can write the energy density with baryon
number density $n$ and proton fraction $x$ as
\beq
\label{eq:epsilon_nxy}
\epsilon(n,x,y) = \epsilon_N(n,x,y) + 
\frac{(3\pi^2 y n)^{5/3}}{10\pi^2 M_\Lambda} + 
y n \left\{  M_\Lambda + \left[E_\Lambda + 
\left( x-\frac{1}{2} \right)^2 S_\Lambda \right]f(n) \right\},
\eeq  
where $\epsilon_N$ is the energy of the nucleons, $E_\Lambda$ is the
shift in the $\Lambda$ energy in nuclear matter (at the saturation
density $n_0$), $S_\Lambda$ parametrizes the shift of this energy as
the proton fraction changes (from nuclear matter with $x=1/2$ to
neutron matter with $x=0$). Finally, the function $f(n)$ (with
$f(n_0)=1$) parametrizes the change in the $\Lambda$ energy as the
baryon density is changed. Before we discuss the bound on each of
these parameters, let us comment on the choice of the form in
eq.~\ref{eq:epsilon_nxy}. Terms linear in $x-1/2$ (or higher odd
powers of $x-1/2$) are expected to be very small as they result from
isospin breaking effects. As commented above, we will be interested in
small values of $y$ for which a linear approximation suffices.
Finally, if $\Lambda$s did not interact with nucleons, the parameters
$E_\Lambda$ and $S_\Lambda=0$ would vanish and the $\Lambda$
contribution to the energy density would be given by the free gas term
(the term proportional to $y^{5/3}$) plus the contribution of their
rest mass.

\begin{figure}
  \includegraphics[width=8cm]{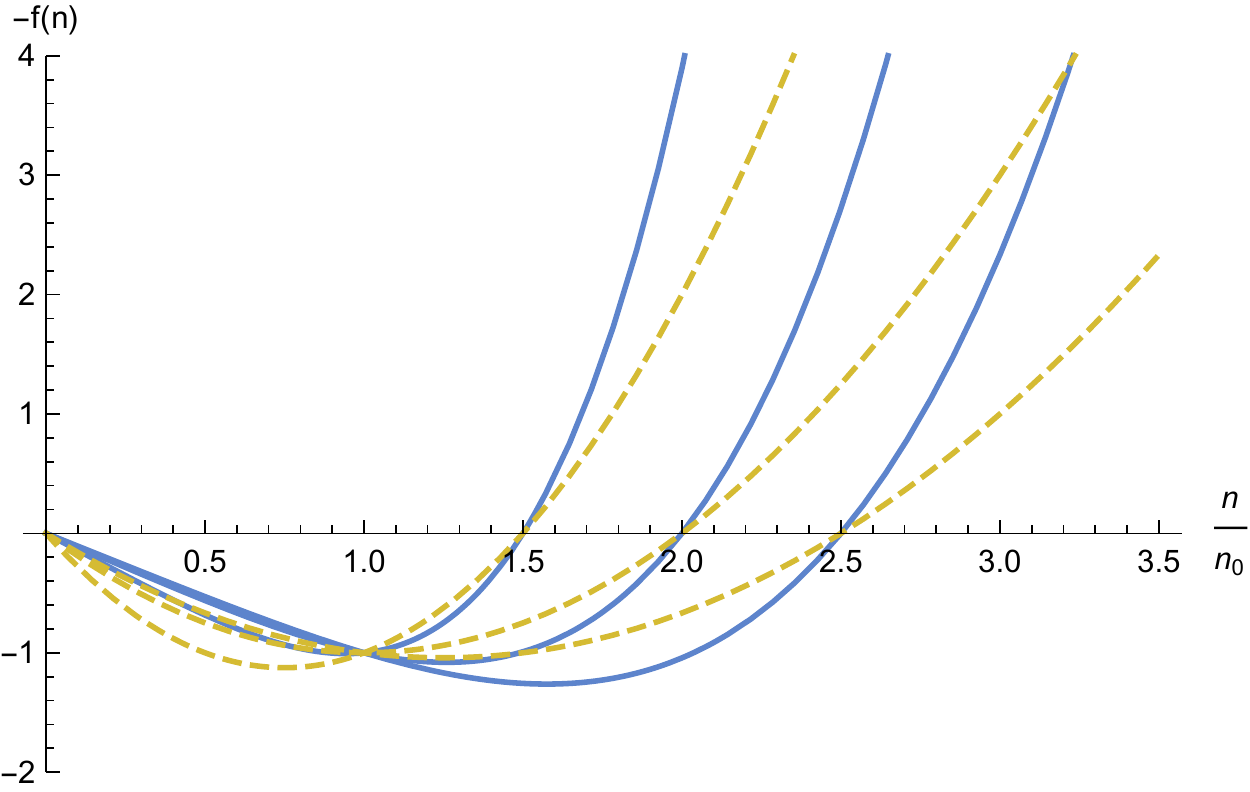}
  \caption{ The function $-f(n/n_0)$ for $\Delta=1.5, 2.0$ and $2.5$
    and $\delta=3$ (solid line, blue online) and $\delta=1$ (dashed,
    yellow online).}
  \label{fig:f}
\end{figure}

The information obtained from the study of hypernuclei discussed above
sets bounds on the values of $E_\Lambda$ and $S_\Lambda$. The
$\Lambda$ energy shift in nuclear matter ($x=1/2$), determines
$E_\Lambda =\Delta E = 28.5\pm 2.0 $ MeV. This determination is
reliable and consistent with the energy shift arising from somewhat
different hypernuclei models (for instance, from models including a
spin-orbit force~\cite{PhysRevC.38.1322}). The bounds on $S_\Lambda$
are looser and follows from our analysis of the previous section
\beq
S_\Lambda=4 d \approx 20 \pm 60 {\rm \ MeV}.
\eeq
Unfortunately, it does not seem to be currently possible to
extract any information about the dependence of the $\Lambda$ energy
away from the saturation density from the phenomenology of
hypernuclei. In our parametrization of the equation of state, this
dependence is contained in the function $f(n)$ and we will use the
existence of two solar mass neutron stars to constrain it. One thing
we do know about $f(n)$ is that $f(n) \sim n$ for small values of $n$
(the proportionality constant being related to the forward scattering
amplitude~\cite{Eletsky:1996jg}). At higher densities, however, the
trend of decreasing in-medium $\Lambda$ mass with the density has to
stop. Otherwise, the number of $\Lambda$s grows quickly and the
equation of state is too soft to support two solar mass stars. Since
we know little about the process leading to the reversal of the trend
we will simply parametrize $f(n)$ in terms of two parameters $\Delta$
and $\delta$:
\beq
f(n/n_0) = \frac{n}{n_0}\frac{1}{1-\Delta^{-\delta}} 
\left[ 1-\left( \frac{n}{\Delta n_0} \right)^\delta\right].
\eeq 
The parameters $\Delta$ fixes the density beyond which the $\Lambda$
in-medium mass is larger than in the vacuum and it is essential for
the plausibility of the equation of state while the parameter $\delta$
determines the shape of the mass dependence with density and is of
lesser importance. The function $f(n/n_0)$ is plotted in
fig.~\ref{fig:f} for several values of $\Delta$ and $\delta$.

Now we will discuss the non-strange equation of state
$\epsilon_N(n,x,y)$. The highest momentum collisions in neutron matter
occur between two back-to-back neutrons at the top of the Fermi
surface. At low densities, below $n\alt 2 n_0$, these momenta are
below the inelastic threshold and non-relativistic potential models
are adequate to describe them. The force between nucleons can be
inferred from the well know measured phase shifts. This program has
been carried out and explain well not only nucleon-nucleon scattering
but, when phenomenological three-body forces are included, the binding
energy of light nuclei. Models with two and three-body forces
determined this way were used to study neutron matter (but not nuclear
matter)~\cite{Gandolfi:2011xu,Steiner:2012,Gandolfi:2013baa} using
Monte Carlo methods. The dependence on the three-body force, which is
less well known than the two-body force, is modest at low densities
but becomes more important at higher densities ($n \agt 2 n_0)$. The
same nucleon-nucleon phase shifts can be fit by potentials obtained
from a low energy chiral expansion. It has been claimed that the
chiral potential, after being evolved according to the renormalization
group, can be used perturbatively in calculations of cold neutron
matter~\cite{Hebeler:2009iv}. The resulting equation of state is very
similar, including its uncertainties, as the one obtained by Monte
Carlo methods~\cite{Hebeler:2010jx, Hebeler:2013nza}. For neutron star
applications it is necessary to perform a small extrapolation of the
pure neutron matter equation of state to non-zero proton fractions
$x\neq 0$. This extrapolation, besides being small ($x \alt 6\%$ near
saturation), is guided by the empirical values of the symmetry energy
and its density dependence. The easiest way to incorporate this
information is to parametrize $\epsilon_N$, for instance, in the
Skyrme-like form:~\cite{Skyrme:1959,Hebeler:2013nza}:
\bea
\label{eq:epsilon}
\epsilon_N(n,x,y) &=&(1-y)n  M_N +
\frac{3n T_0}{5} \left( x^{5/3} + (1-x-y)^{5/3}\right)
\left( \frac{2n}{n_0}\right)^{2/3} \nonumber \\
&-& T_0\left[ (2\alpha-4\alpha_L)x(1-x-y) + 
\alpha_L(1-y)^2 \right]\frac{(1-y)^2n^2}{n_0}\nn\\
&+&(1-y)n T_0\left[ (2\eta-4\eta_L)x(1-x-y) + \eta_L (1-y)^2\right]
\left(\frac{(1-y)n}{n_0}\right)^\gamma,
\eea 
with $T_0=(3\pi^2 n_0/2)^{2/3}/2M_N$. When reduced to pure neutron
matter ($x=y=0$), eq.~\ref{eq:epsilon} fits the Monte Carlo results of
refs.
\cite{Hebeler:2010jx,Hebeler:2013nza,Gandolfi:2011xu,Gandolfi:2013baa}
very well and it is a convenient manner to parametrize them. Away from
$x=0$ it is the most general even function of $x-(1-y)/2$ (as required
by isospin symmetry) up to quadratic order in $x-(1-y)/2$.
    
The five parameters $\alpha, \alpha_L,\eta,\eta_L$ and $\gamma$ can be
determined by the empirical knowledge of five quantities:
\bea
\label{eq:empirical}
-B &=& \epsilon_N(n_0,1/2,0)+\frac{M_N+M_P}{2},\nn\\
p&=&n^2 \frac{\partial \epsilon_N/n}{\partial n}|_{n=n_0, x=1/2,y=0}=0,\nn\\
K &=& 9 n_0^2 \frac{\partial^2\epsilon_N}{\partial n^2}|_{n=n_0, x=1/2, y=0},\nn\\
S &=& \frac{n_0}{8}\frac{\partial^2\epsilon_N}{\partial x^2}|_{n=n_0, x=1/2,y=0},\nn\\
L &=& \frac{3n_0}{8}\frac{\partial^3\epsilon_N}{\partial n\partial x^2}|_{n=n_0, x=1/2,y=0}.
\eea 

The analysis of nuclear masses predicts $B=16\pm 0.1$ MeV and $n_0 =
0.16\pm0.01~\mathrm{fm}^{-3}$~\cite{Kortelainen14} and the study of
giant resonances imply $K=235\pm 25$ MeV for the nuclear
incompressibility. Finally, a wide range of experimental data from
nuclear masses, dipole polarizabilities, and giant resonances implies
$S=32 \pm 2$ MeV for the symmetry energy and $L=50 \pm 15$ MeV (see
\cite{Lattimer:2012xj,Lattimer14a} and references therein). Given
values of $B$, $n_0$ and $K$, one can determine $\alpha$, $\eta$, and
$\gamma$, and then $S$ and $L$ can be used to obtain $\alpha_L$ and
$\eta_L$. 

\begin{figure}
  \includegraphics[width=10cm]{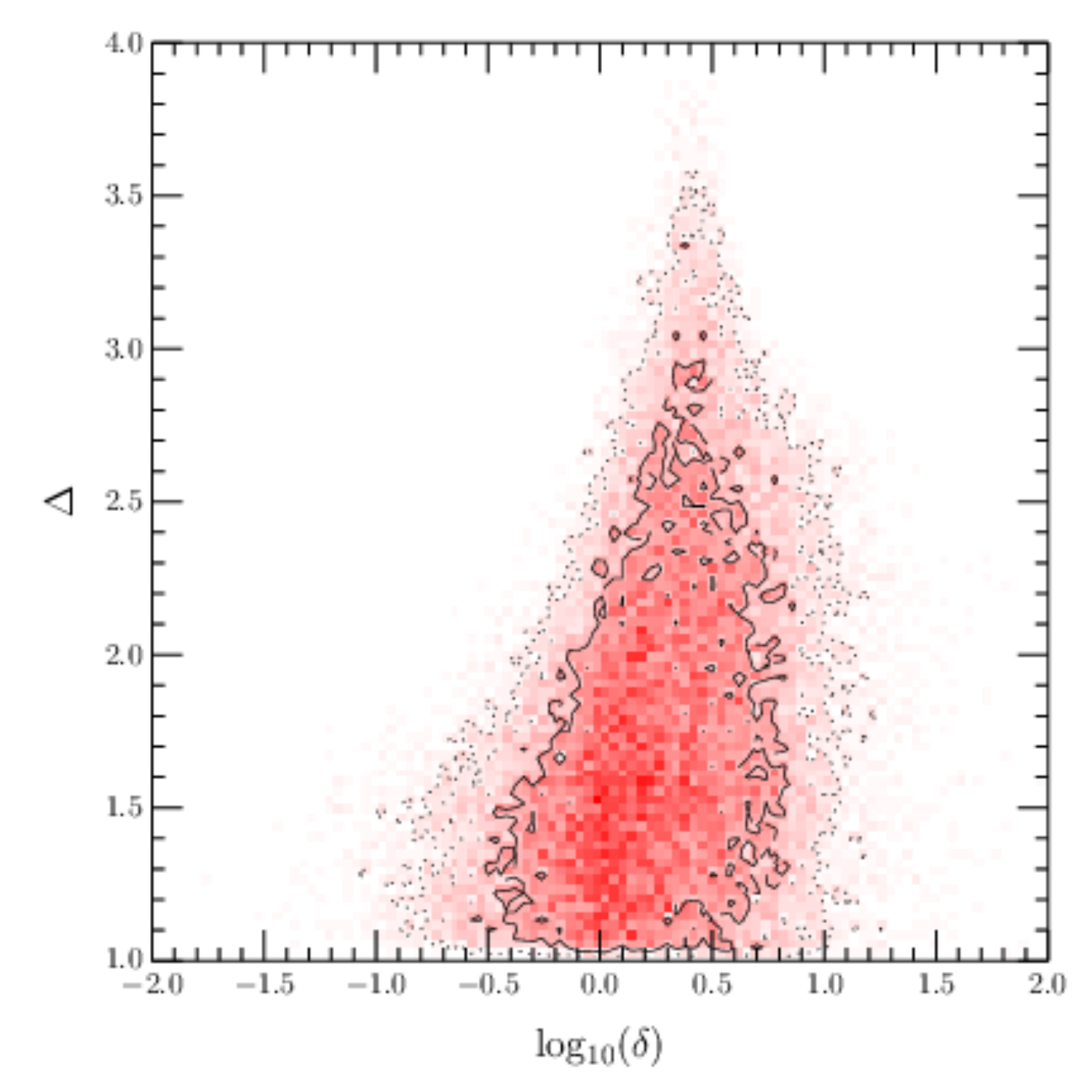}
\noindent
\caption{Monte Carlo probability distribution of neutron
star models with $M > 2 \mathrm{M}_{\odot}$ in the 
$\delta-\Delta$ plane.}
\label{fig:delldel}
\end{figure}  

After a set of parameters is chosen, the $\beta-$equilibrated state is
found by minimizing $\epsilon(n,x,y)$ in relation to $x$ and $y$ for
any given value of $n$. At the highest density considered and for all
parameters used $x<25\%$, confirming that the extrapolation is
relatively small. In any viable equation of state $y$ is very rarely
larger than 10\% substantiating the assumed linear dependence on $y$
on eq.~\ref{eq:epsilon_nxy}.

It follows the discussion above that all but two the parameters in our
equation of state (eq.~\ref{eq:epsilon_nxy}), namely, $\alpha,
\alpha_L, \eta, \eta_L, \gamma, E_\Lambda, S_\Lambda$, are
constrained, to a larger or lesser degree, by empirical information.
We can now use the existence of neutron stars with $M\approx 2
\mathrm{M}_{\odot}$ to constrain the remaining two, $\Delta$ and
$\delta$. The parameter $\Delta$ sets the density, in units of 
the saturation density, beyond which the
$\Lambda$ is repelled, as opposed to attracted, to dense matter. If
$\Delta$ is too large, there is a large range of densities where the
presence of $\Lambda$ particles is energetically favored, making the
equation of state to soft too support $M\approx 2 \mathrm{M}_{\odot}$
stars. From microphysics, the only thing we know about $\Delta$ is
that $\Delta>1$. The parameters $\delta$ fixes the shape of the
in-medium $\Lambda$ energy as a function of the density. As we will
see, neutron star masses puts very loose bounds on it and $\delta$
plays very little role in our discussion.

Taking $B=16$ MeV, $n_0 = 0.16~\mathrm{fm}^{-3}$, we perform a Monte
Carlo simulation (based on the work in
Refs.~\cite{Steiner10,SLB13,Steiner14}) over the estimated values of
the parameters $K$, $S$, $L$ as in Ref.~\cite{Bedaque15} and also over
the new parameters $E_{\Lambda}$, $S_{\Lambda}$, $\Delta$ and
$\delta$. We use Gaussian distributions for all parameters except for
$\Delta$ and $\delta$. For $\Delta$, we use a uniform distribution
between 1 and 10 and for $\delta$ we use a log-normal distribution
centered at 1. The width is fixed at 1/2 (taking logs with base 10)
and the one-sigma range is approximately $1/3$ to $3$. We compute the
equation of state using \ref{eq:epsilon_nxy} and solve the
Tolman-Oppenheimer-Volkov equations, rejecting all points which do not
have $M > 2 \mathrm{M}_{\odot}$. The results are given in
fig.~\ref{fig:delldel}. As $\Delta$ becomes large, the pressure of
matter decreases with increasing energy density, so values of
$\Delta>4$ are not explored by the Monte Carlo. This puts a strict, if
not unsurprising, bound on how fast the $\Lambda$ binding has to
change with density. The upper left corner of the plot is ruled out
(small values of $\delta$ and large values of $\Delta$) by the maximum
mass constraint. This result is consistent with many other previous
works which found that the ``hyperon problem'' could be solved with an
appropriate variation of model parameters. We find this is also the
case in our phenomenological model for small $\Delta$ or large
$\delta$.

\begin{figure}
  \includegraphics[width=10cm]{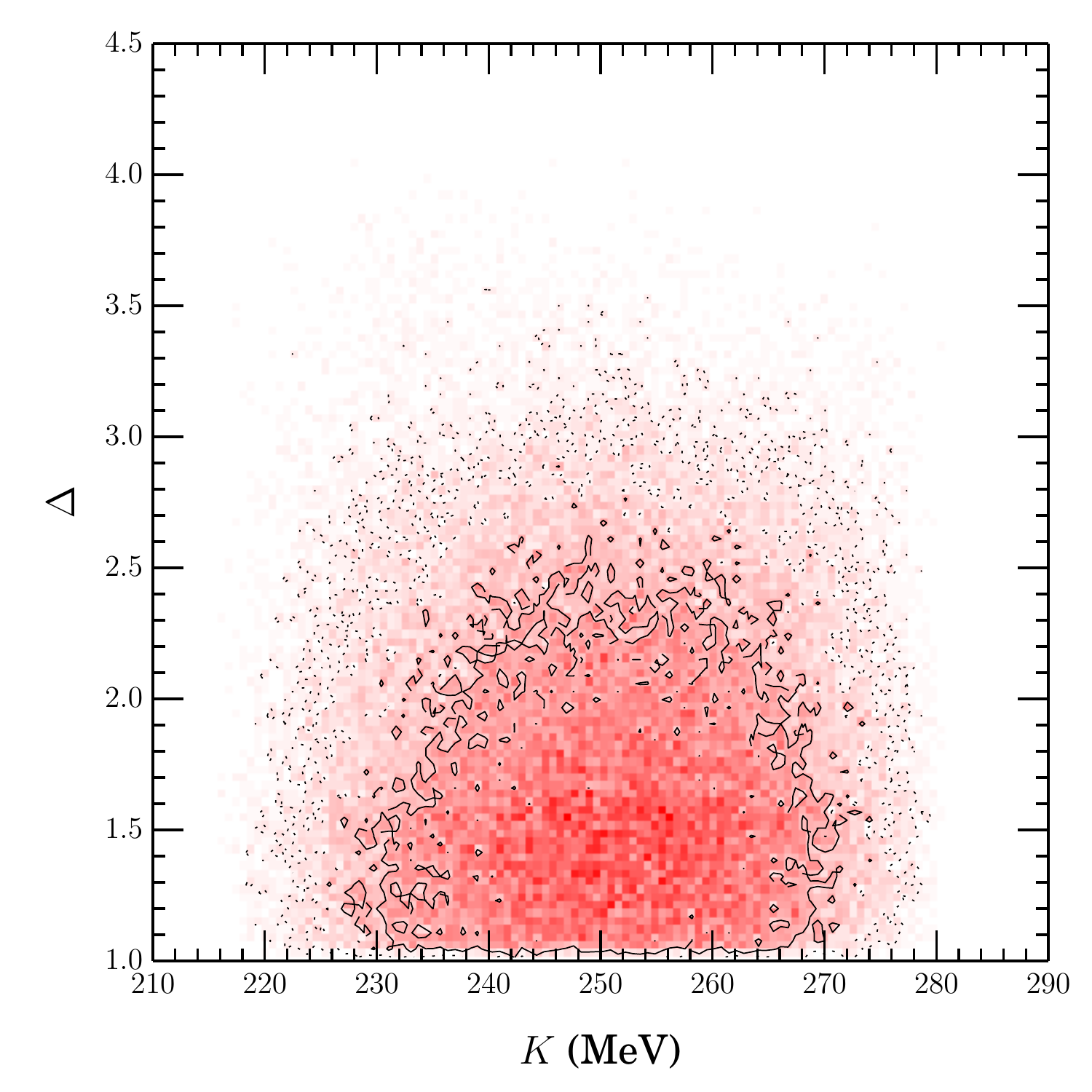}
\noindent
\caption{Monte Carlo probability distribution of incompressbility
  (x-axis) and the parameter $\Delta$ (y-axis) for star models with $M
  > 2 \mathrm{M}_{\odot}$}
\label{fig:KDelta}
\end{figure}

The results above were obtained under some assumptions that we will
now discuss. The first was the non-strange equation of state
eq.~\ref{eq:epsilon}; while it parametrizes well the fairly well know
low density ($n \alt 2 n_0$) equation of state, it is unknown how well
it does at higher density. The high density behavior of
eq.~\ref{eq:epsilon} is dominated by the value of the exponent
$\gamma$: higher $\gamma$ corresponding to stiffer equations of state.
The exponent $\gamma$ itself is largely determined by the value of the
nuclear matter compressibility $K$ and is little affected by the
remaining parameters in eq.~\ref{eq:empirical}. Thus, the non-strange
equations of state with larger values of $K$ are the ones stiffer at
high densities. It turns out, however, that increasing the value of
$K$ -- even outside the empirically acceptable range -- does not
increase the range of value of $\Delta$ consistent with $M > M_\odot$
stars. Fig.~\ref{fig:KDelta} demonstrates this, where it is clear that
$K$ and $\Delta$ are essentially uncorrelated. The reason is that, as
mentioned above, increase the energetic cost of nucleons only triggers
the formation of more $\Lambda$s without an increase of pressure on
energy density. It should be pointed out, however, that with the
assumed form of the $\Lambda$ energy density
(eq.~\ref{eq:epsilon_nxy}), very stiff equations of state lead to an
instability where a large number of $\Lambda$s appear. In a more
complete model including $\Lambda-\Lambda$ interactions this
instability would be cutoff by the repulsion between $\Lambda$s. Thus,
the possibility remains that a stiff non-strange equation of state
coupled to a non-linear dependence of the energy density with the
$\Lambda$ fraction ($y$) would support $M>2.4M_\odot$ stars,
regardless of the value of $\Delta$. The few doubly strange
hypernuclei presently known might be used to extract some information
about $\Lambda-\Lambda$ interaction in nuclear matter but the
extrapolation of that to neutron matter at higher densities seem too
unconstrained to be pursued at the moment. Another possibility is the
onset of quark matter at relatively low
densities~\cite{Chen11,Dexheimer13,Whittenbury14}, maybe even at lower
densities than the threshold for hyperon formation. That possibility,
of course, obviates the hyperon problem while, at the same time,
leading to problems of their own that could be solved by postulating a
stiffer quark matter equation of state~\cite{Blaschke14}.

\section{Conclusion}
  
We argued that the empirical knowledge of the $\Lambda$ separation
energy in $\Lambda$-nuclei provides information not only about the
mass shift of the $\Lambda$ particle in nuclear matter but also some
bounds on the mass shift in {\it neutron} matter. We use this bound to
construct a phenomenological model for the energy density of
neutron/proton/$\Lambda$/electron dense matter. This model contains,
besides the parameters constrained by microphysics, two new parameters
describing the change of the $\Lambda$ mass shift with density. One of
them, $\Delta$, determines the density at which the $\Lambda$-neutron
matter interaction changes from atractive to repulsive. We found that
$\Delta$ is constrained to lie in the range $1<\Delta\alt 3$ in order
to ensure hydrodynamic stability and to ensure that
stars with $M > 2M_\odot$ are supported. We do not find strong
correlations between our model parameters and the radius of low-mass
neutron stars. This implies that other works which have found that
neutron star radius measurements are strong probes of
$\Lambda$-nucleon interactions are at least somewhat model-dependent.
The possible presence of quarks will only strengthen this conclusion.

\begin{acknowledgments}
This material is based upon work supported by the U.S. Department of
Energy Office of Science, Office of Nuclear Physics under Award Number
DE-FG02-93ER-40762. The solver for the stellar structure equations was
obtained from O$_2$scl~\cite{o2scl} and the Monte Carlo was based on the
Bayesian analysis routines in~\cite{bamr}.
\end{acknowledgments}

\bibliographystyle{unsrt}
\bibliography{hypernuclei,velocity_bound} 

\end{document}